\baselineskip=18pt
\voffset=-1.45truecm
\hoffset=.15truecm
\font\ninerm=cmr9

\font\ninebx=cmbx9
\font\nineti=cmti9

\font\eightrm=cmr8
\def\hs{\hskip}

\def\b#1{{\bf #1}}
\def\r#1{{\rm #1}}
\def\h{\hat}
\def\s#1{{^{*}#1}}
\def\d#1{{\delta #1}}
\line{October 1997\hfil ENSK-ph/97-02}
\bigskip\bigskip
\centerline{\bf DAMPING RATE FOR TRANSVERSE GLUONS WITH} 
\centerline{\bf FINITE SOFT MOMENTUM IN HOT QCD}
\bigskip\bigskip
\centerline{A. Abada$^{1}$, O. Azi$^{1}$, and A. Tadji$^{2}$}
\medskip
\centerline{\it $^{1}$ D\'epartement de Physique, \'Ecole Normale
                Sup\'erieure}
\centerline{\it BP 92, Vieux Kouba 16050, Alger, Algeria}
\smallskip
\centerline{\it $^{2}$ D\'epartement de Physique, Universit\'e de Blida}
\centerline{\it BP 270 Blida 09000, Algeria}
\bigskip
\centerline{\tt enskppps@ist.cerist.dz}
\bigskip\bigskip\bigskip
\centerline{\ninebx Abstract}
\bigskip
\hbox{\hs1truein\vbox{\hsize=4.5truein\baselineskip=12pt\ninerm
We calculate the damping rate for transverse gluons with {\nineti finite} 
soft momentum to leading order in perturbative hot QCD. The internal momenta
of the one-loop contributing diagrams are soft. This means we have to use
effective vertices and propagators which incorporate the so-called hard
thermal loops. We expand the damping rate in powers of the incoming momentum
and argue that the series ought to converge within a finite radius of
convergence. We contrast such a behavior with the one obtained from a 
previous calculation that produced a logarithmic behavior, a calculation
based on letting the gluon momentum come from the hard limit down towards the
interior of the soft region. This difference in behavior may point to
interesting physics around some `critical' region.}}
\vfil\eject

\line{\bf 1. Introduction\hfil}
\medskip
The quark-gluon plasma (qgp) is a phase of hot hadronic matter that we hope
to see in very near-future  experiments like RHIC and/or LHC.
One paramount importance of this plasma is that it allows us to see quarks
and gluons, if not completely free, at least in a deconfined plasma phase,
and get clues on the mechanism of confinement. In this respect, many
properties of QCD at (high) finite temperature $T$ have been
investigated [1].

One important quantity that is related to the stability of the qgp
is the damping rate $\gamma(p)$ of gluons\footnote*{\eightrm
We discuss electric gluons only. The interesting magnetic
sector is more intricate as it is well known.}
with momentum $\b p$ in the presumed plasma. This quantity has raised a
great deal of controversy in the past because, when calculated at $\b p=0$
in a standard loop expansion at finite $T$, it is plagued with gauge
dependence both in magnitude and sign [2]. This problem has been solved
when recognizing that in such conditions, the loop expansion is not
necessarily an expansion in powers of the QCD coupling constant $g$,
and hence one has to reorganize the perturbative
expansion so that one takes into account the resummation of the so-called
hard thermal loops, which are loop diagrams with hard (i.e., $\sim T$)
internal momenta [3]. One then argues that $\gamma(0)$ is finite,
gauge-independent and positive.

However, a previous calculation [4] suggests that when we let $p$ run from
the hard limit down towards the interior of the soft region, the boundary of
which being physically set by $m_g$, the inverse thermal gluonic correlation
length or thermal gluonic mass for short (which is of order $gT$),
the damping rate $\gamma_{t}(p)$ for transverse gluons gets a logarithmic
behavior $\ln(1/g)$. This result is in contrast with the fact that
$\gamma_{t}(p=0)$ is finite. It is also in contrast  with the expectation
that the qgp ought to be stable for at least very small but nonzero momenta.
A similar interest was raised in [5] and a discussion in the context of scalar
QED was carried.

In this work, we undertake the expansion in the soft region
$p\le m_g$ of the damping rate $\gamma_{t}(p)$ in powers of $p/m_g$, i.e.,
we write:
$$
 \gamma_{t}(p)= {g^2N_cT\over 24\pi}\,\Big[a_{t0}+a_{t1}\Big
 ({p\over m_g}\Big)^2+ a_{t2}\Big({p\over m_g}\Big)^4+\dots\Big]
 \ ,\eqno(1.1)
$$
where $N_c$ is the number of colors. The quantity ${g^2N_cT\over 24\pi}
a_{t0}$ is just $\gamma_{t}(0)$ with $a_{t0}=+6.63538\dots$, determined
in [2]. Our primary aim is to argue that such an expansion is valid within
a {\it finite} radius of convergence that we denote by $\mu$. We do this by
explicitly calculating the second coefficient $a_{t1}$ in the above expansion
and suggesting that the other coefficients may be calculated in a similar
manner. If indeed this expansion is a valid one, it would mean that the
analytic behavior of the damping rate changes when we cross from the region
below $\mu$ to the region above. Such a result would suggest that there
could be interesting physics to investigate in the `critical' region
around $\mu$. 

This paper is organized as follows. After this introduction,
we set up the stage in the next section for the calculation of the
transverse-gluon damping rate. In the third section, we carry the calculation
of the effective self-enegy to order $p^2$, which is the essential new
quantity entering the definition of $\gamma_t(p)$, see equation (2.9)
below. We discuss our results in the last section and finish with few
concluding remarks.
\bigskip
\line{\bf 2. Preliminaries\hfil}
\medskip
In this section, we prepare the ground for the calculation of the
transverse gluonic damping rate $\gamma_t(p)$.
We work in the imaginary-time formalism in which the euclidean momentum
of the gluon is $P^\mu = (p_0 , \b p)$ such that $P^2 = (p_0 )^2 + p^2 $
where $\b p =p\; \b{\h{p}}$ and $p_0 = 2\pi nT$ where $n$ is an integer.
After we perform the intermediary steps, we obtain the real-time amplitudes
via the analytic continuation $p_0 = -i\omega + 0^+ $ where $\omega$ is
the energy of the gluon. The convention we adopt is that a momentum is said
to be soft if both $\omega$ and $p$  are of order $gT$;
it is said to be hard if one is or both are of order $T$ [3].

We carry our calculation in the Coulomb gauge in which the complete inverse
gluon propagator is given by:
$$
 D_{\mu\nu}^{-1}(P) = P^2 \delta^{\mu\nu} - P^{\mu}P^{\nu}-\Pi^{\mu\nu}(P)
 + {1\over \xi_{C}} \delta^{\mu i}\delta^{\nu j} p^{i}p^{j}\ , \eqno(2.1)
$$
where $\Pi^{\mu\nu}(P)$ is the gluon self-energy and the last term is due to
Coulomb-gauge fixing. In fact, it is most suitable to work in the strict
Coulomb gauge $\xi_{C}=0$. The gluon self-energy can be decomposed into:
$$
 \Pi^{\mu\nu}(P)=\d{\Pi}^{\mu\nu}(P) + \s{\Pi}^{\mu\nu}(P)\ , \eqno(2.2)
$$
where $\d{\Pi}$ is the hard thermal loop and  $\s{\Pi}$ is the effective
self-energy. $P$ being soft, the hard thermal loop is of the same order of
magnitude as the inverse free propagator, i.e., $\d{\Pi}\sim (gT)^2$, while
the effective self-energy is of an order of magnitude higher, i.e., 
$\s{\Pi}\sim g(gT)^2$. Since the momentum running inside  $\s{\Pi}$ is soft,
we have to use effective vertices and propagators instead of their
bare\footnote{*}{\eightrm `Bare' refers here to the usual quantities one
considers as dictated by the Feynman rules.} counterparts when calculating it.
This ensures the correct expression for the $g(gT)^2$--correction to the
inverse gluon propagator and, in particular, that this correction is
independent of the gauge.

The hard thermal loop $\delta\Pi$ can already be found in the literature,
see for example [2,3].
It is real and contributes to the determination of the spectrum of
the soft gluonic exitations to leading order $gT$.
More explicitly, we know that $\d{\Pi}$ is gauge-invariant and  
satisfies the identity $P^{\mu}\d{\Pi}^{\mu\nu}(P)=0$. 
This means its components can be expressed in terms of only two 
independent scalar functions denoted by $\d{\Pi}_{l}(P)$ and
$\d{\Pi}_{t}(P)$ such that:
$$
 \eqalign{\d\Pi^{00}(P)& = \d\Pi_{l}(P)\ ;\qquad 
 \d\Pi^{0i}(P) = - {p_0 p^i\over p^2}\ \d\Pi_{l}(P)\ ;\cr
 \d\Pi^{ij}(P) &= (\delta^{ij}-\h{p}^i \h{p}^j)\ \d{\Pi}_{t}(P)
 + \h{p}^i \h{p}^j\ {(p_0)^2\over p^2}\ \d\Pi_{l}(P)\ .\cr}\eqno(2.3)
$$
The expressions of $\delta\Pi_{l}(P)$ and $\delta\Pi_{t}(P)$ 
read [3]:
$$
 \d\Pi_{l}(P)=3 m_{g}^2\ Q_{1}\Big({ip_0\over p}\Big)\ ;\qquad
 \d{\Pi}_{t}(P) = {3\over 5}\,m_{g}^2\,\Big[Q_{3}\Big({ip_0\over p}\Big)
 - Q_{1}\Big({ip_0 \over p}\Big) - {5\over 3}\Big]\ ,
 \eqno(2.4)
$$
where the $Q_n$ are Legendre functions of the second kind. As already
mentioned, $m_g$ is the gluon thermal mass and, to lowest order, is equal to 
$\sqrt{N_c+(1/2)\,N_f}\,gT/3$, where $N_f$ is the number of flavors.

The effective propagator for soft gluons that intervenes in the calculation
of the effective self-energy is obtained by inverting (2.1) while
disregarding $\s\Pi$. In the strict Coulomb gauge, its nonzero components
are $\s{\Delta}_{C}^{00}(P)= \s{\Delta}_{l}(P)$ and
$\s{\Delta}_{C}^{ij}(P) = (\delta^{ij}-\h{p}^i\h{p}^j)
\hskip2pt\s{\Delta}_{t}(P)$, where $\s{\Delta}_{t}$ 
and $\s{\Delta}_{l}$ are given by:
$$
 \s{\Delta}_{t}(P) = {1\over{P^2 - \d{\Pi}_{t}(P)}}\ ; \qquad 
 \s{\Delta}_{l}(P) = {1\over{p^2 - \d{\Pi}_{l}(P)}}\ .    \eqno(2.5)
$$
After analytic continuation to real energies, the pole in $\omega$ 
of $\s\Delta_{t(l)}$ yields the dispersion relation $\omega_{t(l)}(p)$ 
for the transverse (longitudinal) gluons to order $gT$. One finds for
the soft transverse ones:
$$
 \omega_t(p)= m_g\,\Big[1+ {3\over 5}\Big({p\over m_g}\Big)^2 
  - {9\over 35}\Big({p\over m_g}\Big)^{4} +{704\over 3000}
  \Big({p\over m_g}\Big)^{6}-{91617\over 336875}
  \Big({p\over m_g}\Big)^{8} + \dots\Big]\ .\eqno(2.6)
$$

As mentioned earlier, the hard thermal loop $\delta\Pi$ is
real above the light cone, and so the poles of (2.5) are real, which
means the gluons are not damped to this order $gT$; this is clear from
(2.6). In order to get the
leading order of the damping rates, we have to include in the dispersion
equations the contribution from the effective self-energy. $\s\Pi$ has of
course a more complicated structure than $\delta\Pi$. It satisfies the less
restrictive identity $P^\mu\ \s{\Pi}^{\mu\nu}(P)\ P^\nu = 0$ [3]. This means
that in general, $\s{\Pi}^{\mu\nu}(P)$ can be written in terms of three
independent scalar functions, but in the strict Coulomb gauge, only two
of these are relevant. The transverse dispersion relation including the
self energy reads:
$$
 -\Omega_t^2 + p^2 - \d\Pi_t(-i\Omega_t,p)
 - \s\Pi_t(-i\Omega_t,p) = 0\ ,\eqno(2.7)
$$
where $\s\Pi_t$ is given by:
$$
 \s\Pi_t(P)={1\over 2}(\delta^{ij}-\h p^i\h p^j)\ \s\Pi^{ij}(P)\ .
 \eqno(2.8)
$$
The transverse gluon damping rate is defined by $\gamma_t(p) \equiv
-{\rm Im}\ \Omega_t(p)$. Since it is $g$-times smaller than
the energy $\omega_t(p)$, we can write from (2.7):
$$
 \gamma_t(p) = {{\rm Im}\s\Pi_t(-i\omega, p)\over
 2\omega+{\partial\over\partial\omega}
 \d\Pi_t(-i\omega, p)}\ \Big{|}_{\omega=\omega_t(p)+i0^+}\ .\eqno(2.9)
$$
The denominator in (2.9) is easy to get since we already have an expression
for the hard thermal loop $\d\Pi_t$. Indeed, we have:
$$
 2\omega_t(p)+\partial_{\omega}\d\Pi_t(-i\omega,p)|_{\omega=\omega_t(p)
 +i0^+} = 2\, \big[1+{1\over 10}\big({p\over m_g}\big)^2+\dots\big]\ .
 \eqno(2.10)
$$
This means our main task is to
calculate the imaginary part of $\s\Pi_t$. This we do in the sequel.
\bigskip
\line{\bf 3. The imaginary part of the transverse effective
          self-energy\hfil}
\medskip
In the Coulomb gauge, the only diagrams that contribute to the imaginary
part of the effective self-energy above the light cone are the three-gluon
and four-gluon one loop-diagrams with soft internal momentum [3].
Hence we write:
$$
 \eqalignno{{\rm Im}\ \s\Pi^{\mu\nu}(P)= &
 -{g^{2}N_{c}\over 2}\ {\rm Im\ Tr_{soft}}\
 \big[\s\Gamma^{\mu\nu\lambda\sigma}(P,-P,K,-K)
 \ \s\Delta_{C}^{\lambda\sigma}(K)\cr & +
 \s\Gamma^{\sigma\mu\lambda}(-Q,P,-K)
 \ \s\Delta_{C}^{\lambda\lambda'}(K) 
 \ \s\Gamma^{\lambda'\nu\sigma'}(-K,P,-Q) 
 \ \s\Delta_{C}^{\sigma'\sigma}(Q)\big],   &(3.1)\cr}
$$
where $K$ is the internal loop-momentum, $Q=P-K$ and
Tr $\equiv T\sum_{k_0}\int {d^{3}k\over (2\pi)^3}$\hs2pt. The subscript
`soft' means that only soft values of $k$ are allowed in the integral.
Eq (3.1) is what one would normally write for the three-gluon and
four-gluon contributions to the imaginary part of the gluon self-energy, 
except that everywhere, bare quantities are replaced by the corresponding 
effective ones.

We have already given in the last section the expressions of the nonvanishing
components of the effective propagators (see eq (2.5) and the text before it).
The gluon effective vertices can be written as:
$$
 \s\Gamma^{(n)} = \Gamma^{(n)} +\ \d\Gamma^{(n)}\ ; \qquad
 n=3,4\,, \eqno(3.2)
$$
where the first term is the QCD gluon tree vertex and the second one sums up
the contributions from hard thermal loops with $n$ external legs. In the case
$n=3$ it can be written as:
$$
 \d\Gamma^{\mu\nu\lambda}(-Q,P,-K) = -\d\Gamma^{\mu\nu\lambda}(-K,P,-Q)=
 3m_{g}^2 \int{d\Omega_S\over 4\pi}
 {{S^\mu S^\nu S^\lambda}\over PS} \Big({iq_0\over QS} 
 -{ik_0\over KS}\Big)\ ,\eqno(3.3)
$$
where $S\equiv (i, \h\b s)$ and $\Omega_S$ is the solid angle of the unit
vector $\h\b s$. Also, $PS=ip_0+ \b p.\h\b s$, etc. In the case
$n=4$ we have:
$$
 \d\Gamma^{\mu\nu\lambda\sigma}(P,-P,K,-K)= 3 m_g^2
 \int{d\Omega_S\over 4\pi} {{S^{\mu} S^{\nu} S^{\lambda} S^{\sigma}}
 \over{PS\ KS}}\Big({{ip_0-ik_0}\over{PS-KS}}-{{ip_0+ik_0}
 \over{PS+KS}}\Big)\ . \eqno(3.4)
$$
To be complete, we give the expression of the three-gluon tree vertex:
$$
 \eqalign{\Gamma^{\mu\nu\lambda}(-Q,P,-K)&=
 -\Gamma^{\lambda\nu\mu}(-K,P,-Q)\cr
 &= (P+K)^{\mu}\,\delta^{\nu\lambda}+(Q-K)^{\nu}\,
 \delta^{\lambda\mu}-(P+Q)^{\lambda}\,\delta^{\mu\nu}\ , \cr}\eqno(3.5)
$$
and that of the four-gluon tree vertex:
$$
 \Gamma^{\mu\nu\lambda\sigma}(P,-P,K,-K)=
 2\,\delta^{\mu\nu}\delta^{\lambda\sigma} - \delta^{\mu\lambda}
 \delta^{\nu\sigma} - \delta^{\mu\sigma}\delta^{\nu\lambda}\ . \eqno(3.6)
$$

From eqs (2.8) and (3.1) above, we can write more explicitly the expression
of the transverse effective self-energy:
$$
 \eqalign{&{\rm Im}\ \s\Pi_t(P)=-{g^2N_c\over 4}\
 (\delta^{ij}-\h{p}^i \h{p}^j)\,{\rm Im}\,T\sum_{k_0}
 \int{d^3k\over(2\pi)^3}\,\big[\s\Gamma^{ij00}(P,-P,K,-K)\,\s\Delta_l(K)\cr
 &+\s\Gamma^{ijmn}(P,-P,K,-K)\,(\delta^{mn}-\h{k}^m\h{k}^n)\,\s\Delta_t(K)\cr
 &+\s\Gamma^{0i0}(-Q,P,-K)\,\s\Delta_l(K)\,
 \s\Gamma^{0j0}(-K,P,-Q)\,\s\Delta_l(Q)\cr
 &+\s\Gamma^{0im}(-Q,P,-K)\,(\delta^{mn}-\h{k}^m\h{k}^n)\,\s\Delta_t(K)\,
 \s\Gamma^{nj0}(-K,P,-Q)\,\s\Delta_l(Q)\cr
 &+\s\Gamma^{mi0}(-Q,P,-K)\,\s\Delta_l(K)\,
 \s\Gamma^{0jn}(-K,P,-Q)\,(\delta^{nm}-\h{q}^n\h{q}^m)\,\s\Delta_t(Q)\cr
 &+\s\Gamma^{mir}(-Q,P,-K)\,(\delta^{rs}-\h{k}^r\h{k}^s)\,\s\Delta_t(K)\,
 \s\Gamma^{sjn}(-K,P,-Q)\,(\delta^{nm}-\h{q}^n\h{q}^m)\,
 \s\Delta_t(Q)\,\big]\ .\cr}
 \eqno(3.7)
$$
There are six contributions: two from the four-gluon vertices (subscript
$4g$ in the sequel) and four from the three-gluon vertices
(subscript $3g$). Each one has to be calculated separately.
As an illustration, we show how we carry the calculation corresponding
to the contribution from the three-gluon vertices where the two effective
propagators involved are both longitudinal. We denote this contribution by
${\rm Im}\,\s\Pi_{t3gll}(P)$ and the others correspondingly.
The other contributions are manipulated in a similar way,
with varying difficulties that we comment on later in this section.
Also, we will take henceforth $m_g\equiv 1$ and all momenta and energies
are in units of it. This simplifies considerably the expressions we write
down and, if and when needed, the $m_g-$dependence can be recovered in the
final results.

Using the expressions of the gluon vertices we gave in eqs (3.2)-(3.6), we
write:
$$
 \eqalign{\r{Im}\,\s\Pi_{t3gll}(P)=&{g^2N_c\over 8\pi^2}\,\r{Im}\,T\sum_{k_0}
 \int{d^3k\over 4\pi}\,\Big[\big(\b{p}-2\b{k}\big)^2-
 \big(\b{p}-2\b{k}\big)\h\b{p}\,^2\cr&
 +6\int{d\Omega_S\over 4\pi}{(\b{p}-2\b{k})\h\b{s}-
 (\b{p}-2\b{k})\h\b{p}\,\h\b{s}\h\b{p}\over PS}\,\Big({ik_0\over KS}
 -{iq_0\over QS}\Big)\cr&+9\int{d\Omega_{S_1}\over 4\pi}
 \int{d\Omega_{S_2}\over 4\pi}\,{\h\b{s}_1\h\b{s}_2-\h\b{s}_1\h\b{p}\,
 \h\b{s}_2\h\b{p}\over PS_1\,PS_2}\,\Big({ik_0\over KS_1}
 -{iq_0\over QS_1}\Big)\,\Big({ik_0\over KS_2}-{iq_0\over QS_2}\Big)\,\Big]
 \,\s\Delta_l(K)\,\s\Delta_l(Q)\ .\cr}\eqno(3.8)
$$
We first work on the term that does not involve a solid-angle integral.
We use the relation $(\b{p}-2\b{k})^2-(\b{p}-2\b{k})\h\b{p}\,^2 =
4k^2\sin^2\psi$, where $\psi=(\h\b{p},\h\b{k})$, and integrate over the
solid angle of $h\b{k}$ after we expand the effective propagator
at the momentum $Q=P-K$ in the following manner:
$$
 \s\Delta_l(Q)=\Big[1-p\cos\psi\,\partial_k+{p^2\over 2}
 \Big({\sin^2\psi\over k}
 \partial_k+\cos^2\psi\,\partial_k^2\Big)+\dots\Big]\ \s\Delta_l(q_0,k)\ ,
 \eqno(3.9)
$$
where $\partial_k=\partial/\partial k$. We find:
$$
 \eqalign{\r{Im}\,T\sum_{k_0}&\int{d^3k\over 4\pi}\,\big[
 \big(\b{p}-2\b{k}\big)^2-\big(\b{p}-2\b{k}\big)\h\b{p}\,^2\,\big]
 \,\s\Delta_l(K)\,\s\Delta_l(Q)\cr&={8\over 3}\,\r{Im}\,T\sum_{k_0}
 \int_0^{\infty}k^4\,dk\,\s\Delta_l(K)\,\Big[1+{p^2\over 5}\,
 \Big({2\over k}\partial_k+{1\over 2}\partial_k^2\Big)+\dots\Big]\,
 \s\Delta_l(q_0,k)\ .\cr}\eqno(3.10)
$$
The next step in to perform the sum over $k_0$, but we take
care of that a little later.

We leave the expression in
(3.10) as it is for the moment and turn to the term in
(3.8) that involves one solid-angle integral. It is sufficient to
concentrate only on the piece that contains the ratio $ik_0/KS$ because
the other one that contains the ratio $iq_0/QS$ is in fact equal to
the first one. We need an expression for the
solid-angle integral. For this purpose, it is best to mesure the solid angle
$\Omega_S=(\theta,\phi)$ with respect to $\h\b k$ such that
$\theta=(\h\b{k},\h\b{s})$. Also, we expand $1/PS$ in the following manner:
$$
 {1\over PS}={1\over ip_0}\,\Big(1-{\b{p}\h\b{s}\over ip_0}
 -{\b{p}\h\b{s}\,^2\over p_0^2}+\dots\Big)\ ,\eqno(3.11)
$$
an expansion valid as long as $|p/ip_0|<1$, which is satisfied for
soft gluons before and after analytic continuation. With this,
we can write:
$$
 \eqalign{\int{d\Omega_S\over 4\pi}\,&{(\b{p}-2\b{k})\h\b{s}
 -(\b{p}-2\b{k})\h\b{p}\,\h\b{s}\h\b{p}\over PS}\,
 {ik_0\over KS}\cr&=-2k\sin\psi\,{ik_0\over ip_0}\,
 \int{d\Omega_S\over 4\pi}\,
 {\sin\psi\,\cos\theta+\cos\psi\,\sin\theta\,\sin\phi
 \over (ik_0+k\cos\theta)}\,\Big(1-{\b{p}\h\b{s}\over ip_0}
 -{\b{p}\h\b{s}\,^2\over p_0^2}+\dots\Big)\ .\cr}
 \eqno(3.12)
$$
With the relation $\h\b{p}\h\b{s}=\cos\psi\,\cos\theta-\sin\psi\,\sin\theta
\,\sin\phi$, the angular integrals are performed straightforwardly
and we end up with:
$$
 \eqalign{\int{d\Omega_S\over 4\pi}\,{(\b{p}-2\b{k})\h\b{s}
 -(\b{p}-2\b{k})\h\b{p}\,\h\b{s}\h\b{p}\over PS}\,{ik_0\over KS}
 =&-{ik_0\over ip_0}(1-x^2)\,\Big[2\Big(1-{ik_0\over k}Q_{0k}\Big)
 +{p\over ip_0}x\,\Big(3{ik_0\over k}\cr&
 +\Big(1+3{k_0^2\over k^2}\Big)Q_{0k}\Big)
 -{p^2\over p_0^2}\Big[{2\over 3}(1-2x^2)+(1-5x^2){k_0^2\over k^2}\cr&
 -{ik_0\over k}\Big(1-3x^2+(1-5x^2){k_0^2\over k^2}\Big)
 Q_{0k}\Big]+\dots\Big]\ ,\cr}
 \eqno(3.13)
$$
where $x=\cos\psi$ and $Q_{0k}=Q_0\big({ik_0\over k}\big)$. 
We plug this expression back into the integral over $d^3k$ and
use (3.9) to perform the integral over the solid angle of $\h\b{k}$.
That is quite straightforward and we get:
$$
 \eqalign{6\,\r{Im}\,&T\sum_{k_0}\int{d^3k\over 4\pi}
 \int{d\Omega_S\over 4\pi}\,
 {(\b{p}-2\b{k})\h\b{s}-(\b{p}-2\b{k})\h\b{p}\,\h\b{s}\h\b{p}\over PS}
 \,\Big({ik_0\over KS}-{iq_0\over QS}\Big)\,\s\Delta_l(K)\,\s\Delta_l(Q)\cr&
 =-16\,\r{Im}\,T\sum_{k_0}\int_{0}^{\infty}k^2\,dk\,\s\Delta_l(K)\,
 {ik_0\over ip_0}\Big[\Big(1-{ik_0\over k}Q_{0k}\Big)+
 {p^2\over 5p_0^2}\Big[-1+{ik_0\over k}Q_{0k}\cr&+{ip_0\over 2}
 \Big(3{ik_0\over k}+\Big(1+3{k_0^2\over k^2}\Big)Q_{0k}\Big)\,\partial_k
 +p_0^2\Big(1-{ik_0\over k}Q_{0k}\Big)\,\Big({2\over k}\partial_k+{1\over 2}
 \partial_k^2\Big)\Big]+\dots\Big]\,\s\Delta_l(q_0,k)\ .\cr}
 \eqno(3.14)
$$

Here also we leave this expression as it is for the moment and turn our
attention to the term in (3.8) involving two solid-angle integrals.
this term is actually equal to
$$
 -18\,\r{Im}\,T\sum_{k_0}\int
 {d^3k\over 4\pi}\,\s\Delta_l(K)\,\s\Delta_l(Q)\,
 \int{d\Omega_{S1}\over4\pi}\int{d\Omega_{S2}\over 4\pi}
 {\h\b{s}_1\h\b{s}_2-\h\b{s}_1\h\b{p}\,
 \h\b{s}_2\h\b{p}\over PS_1\,PS_2}\Big({k_0^2\over KS_1\,KS_2}-
 {k_0q_0\over KS_1\,QS_2}\Big)\
$$
and we first work out the piece that contains
$k_0^2/KS_1\,KS_2$. In order to be able to carry
forward, it is best to write the double solid-angle integral in the
following manner:
$$
 \eqalign{\int{d\Omega_{S_1}\over 4\pi}\int{d\Omega_{S_2}\over 4\pi}\,
 {\h\b{s}_1\h\b{s}_2-\h\b{s}_1\h\b{p}\,\h\b{s}_2\h\b{p}\over PS_1\,PS_2}\,
 {k_0^2\over KS_1\,KS_2}=&k_0^2\Big[\sin^2\psi\,\Big[
 \int{d\Omega_{S}\over 4\pi}{\cos\theta\over PS\,KS}\Big]^2
 +\cos^2\psi\,\Big[\int{d\Omega_S\over 4\pi}
 {\sin\theta\,\sin\phi\over PS\,KS}\Big]^2
 \cr&+2\cos\psi\,\sin\psi\int{d\Omega_S\over 4\pi}{\cos\theta\over PS\,KS}
 \int{d\Omega_S\over 4\pi}\,{\sin\theta\,\sin\phi\over PS\,KS}\ \Big]\ .
 \cr}\eqno(3.15)
$$
Each single solid-angle integral that is involved in the above
expression can be worked out straightforwardly as before, using the
expansion (3.11). Putting things together, we find:
$$
 \eqalign{\int{d\Omega_{S_1}\over 4\pi}\int{d\Omega_{S_2}\over 4\pi}
 \,{\h\b{s}_1\h\b{s}_2-\h\b{s}_1\h\b{p}\,\h\b{s}_2\h\b{p}\over PS_1\,PS_2}
 \,&{k_0^2\over KS_1\,KS_2}=-{k_0^2(1-x^2)\over p_0^2\,k^2}\Big[
 \Big(1-{ik_0\over k}Q_{0k}\Big)^2+
 \cr&\hs-65pt{px\over ip_0}\Big(1-{ik_0\over k}Q_{0k}
 \Big)\,\Big(3{ik_0\over k}+\Big(1+3{k_0^2\over k^2}\Big)Q_{0k}\Big)
 -{p^2\over p_0^2}\,\Big[{x^2\over 4}\Big({ik_0\over k}+
 \Big(1+{k_0^2\over k^2}\Big)Q_{0k}\Big)^2\cr&\hs-65pt
 +\Big(1-{ik_0\over k}Q_{0k}\Big)\,
 \Big({2\over 3}-{4\over 3}x^2+(1-7x^2){k_0^2\over k^2}-
 \Big(1-4x^2+(1-7x^2){k_0^2\over k^2}\Big){ik_0\over k}Q_{0k}\Big)\,\Big]
 +\dots\Big]\ .\cr}\eqno(3.16)
$$
We put this expression back under $\int d^3k$ and perform the integral over
the solid angle of $\h\b{k}$. We get:
$$
 \eqalign{\r{Im}\,T\sum_{k_0}\int{d^3k\over 4\pi}\,&\s\Delta_l(K)\,
 \s\Delta_l(Q)\int{d\Omega_{S_1}\over 4\pi}\int{d\Omega_{S_2}\over 4\pi}\,
 {\h\b{s}_1\h\b{s}_2-\h\b{s}_1\h\b{p}\,\h\b{s}_2\h\b{p}\over PS_1\,PS_2}
 \,{k_0^2\over KS_1\,KS_2}\cr&
 =-{2\over 3}\,\r{Im}\,T\sum_{k_0}{k_0^2\over p_0^2}\int_{0}^{\infty}dk
 \,\s\Delta_l(K)\,\Big[\Big(1-{ik_0\over k}Q_{0k}\Big)^2
 -{p^2\over 5p_0^2}\Big[{1\over 4}\Big({ik_0\over k}+\Big(1+{k_0^2\over k^2}
 \Big)Q_{0k}\Big)^2\cr&\hs10pt
 +\Big(1-{ik_0\over k}Q_{0k}\Big)\,\Big(2-2{k_0^2\over k^2}
 -\Big(1-2{k_0^2\over k^2}\Big){ik_0\over k}Q_{0k}\Big)
 -ip_0\Big(1-{ik_0\over k}Q_{0k}\Big)\,\cr&\hs10pt\times
 \Big(3{ik_0\over k}+\Big(1+3{k_0^2\over k^2}\Big)Q_{0k}\Big)\,\partial_k
 -{p_0^2\over 2}\Big(1-{ik_0\over k}Q_{0k}\Big)^2\,
 \Big({4\over k}\partial_k
 +\partial_k^2\Big)\,\Big]+\dots\Big]\,\s\Delta_l(q_0,k)\ .\cr}\eqno(3.17)
$$

Next we turn to the piece that contains $k_0q_0/KS_1\,QS_2$. To get a
manageable expression for the double solid-angle integral, it is most
suitable to measure the solid angles with respect to $\h\b{p}$.
Then we have:
$$
 \int{d\Omega_{S_1}\over 4\pi}\int{d\Omega_{S_2}\over 4\pi}\,
 {\h\b{s}_1\h\b{s}_2-\h\b{s}_1\h\b{p}\,\h\b{s}_2\h\b{p}\over PS_1\,PS_2}\,
 {k_0q_0\over KS_1\,QS_2}=k_0q_0\,\int{d\Omega_S\over 4\pi}
 {\sin\theta_0\sin\phi_0\over PS\,KS}\int{d\Omega_S\over 4\pi}
 {\sin\theta_0\sin\phi_0\over PS\,QS}\ ,\eqno(3.18)
$$
where $(\theta_0,\phi_0)$ is the solid angle of $\h\b{s}$ with
respect to $\h\b{p}$. Each solid-angle integral can be carried
through separately. We get the following result:
$$
 \eqalign{\int{d\Omega_S\over 4\pi}{\sin\theta_0\,\sin\phi_0\over
 PS\,KS}=& {\sin\psi\over ip_0k}\Big[1-{ik_0\over k}Q_{0k}+
 {px\over 2ip_0}\Big(3{ik_0\over k}+\Big(1+3{k_0^2\over k^2}\Big)
 Q_{0k}\Big)\cr&-{p^2\over 2p_0^2}\Big[{2\over 3}(1-2x^2)+(1-5x^2)
 {k_0^2\over k^2}-\Big(1-3x^2+(1-5x^2){k_0^2\over k^2}\Big)
 {ik_0\over k}Q_{0k}\Big]\dots\Big]\ ,\cr}\eqno(3.19)
$$
and a similar one for the other integral in (3.18) where one replaces $K$
by $Q$ and the angle $(\h\b p, \h\b k)$ by $(\h\b p, \h\b q)$. 
We then multiply the two obtained expressions and put back the result
under $T\sum_{k_0}\int d^3k/4\pi$. But here the integral over the
solid angle of $\h\b k$ is not straightforward yet because the integrand
still depends on $q$ and the angle $(\h\b p, \h\b q)$. Hence a further
expansion is necessary, but instead of expanding $Q_{0q}$ directly, it is
most suitable to write it in terms of $\s\Delta_l^{-1}(Q)$ using (2.4)
and (2.5) and expand the resulting expression. The calculation carries
thereon straightforwardly and we get:
$$
 \eqalign{18\,\r{Im}\,T\sum_{k_0}\int{d^3k\over 4\pi}& \s\Delta_l(K)\,
 \s\Delta_l(Q)\int{d\Omega_{S_1}\over 4\pi}\int{d\Omega_{S_2}\over 4\pi}
 \,{\h\b s_1 \h\b s_2 - \h\b s_1\h\b p\,\h\b s_2\h\b p\over PS_1\,PS_2}
 \, {k_0q_0\over KS_1\,QS_2}={4\over 3}\,\r{Im}\,T\sum_{k_0}\int_{0}^{\infty}
 k^2 dk\cr&{\s\Delta_l(K)\over -p_0^2}\Big[ik_0iq_0k^2+{p^2\over 5p_0^2}
 \Big[-2ik_0iq_0k^2+{1\over 4}(3+k^2+3k_0^2)\,(3+k^2+3q_0^2)
 \cr&-iq_0ip_0k(3+k^2+3k_0^2)\partial_k+ik_0iq_0p_0^2k^2
 \Big({2\over k}\partial_k+{1\over 2}\partial_k^2\Big)\Big]+\dots\Big]
 \,\s\Delta_l(q_0,k)\ .\cr}\eqno(3.20)
$$

It is time now we get into performing the sum over $k_0$. Because of the
complicated $k_0-$dependence of the expressions we work with, it is best
to perform the sum using the spectral representation of the different
quantities involved [6]. For example, we have for the effective propagators:
$$
 \s\Delta_{t,l}(K)=\int_{0}^{1/T} d\tau\,e^{ik_0\tau}\,
 \int_{-\infty}^{+\infty}d\omega\,\rho_{t,l}(\omega,k)\,
 [1+n(\omega)]\,e^{-\omega\tau}\ ,\eqno(3.21)
$$
where $n(\omega)=1/(\exp(\omega/T)-1)$ is the Bose-Einstein distribution
and the spectral function $\rho_{l,t}$ is given by:
$$
 \rho_{t,l}(\omega,k)={Z_{t,l}(k)\over 2\omega_{t,l}(k)}\,
 \big[\delta\big(\omega-\omega_{t,l}(k)\big)
 -\delta\big(\omega+\omega_{t,l}(k)\big)\big]
 +\beta_{t,l}(\omega,k)\,\Theta\big(k^2-\omega^2\big)\ ,\eqno(3.22)
$$
expression in which the residue $Z_{t,l}(k)$ is given by:
$$
 Z_t(k)={\omega(\omega^2-k^2)\over 3\omega^2-
 (\omega^2-k^2)^2}\Big{|}_{\omega=\omega_t(k)}\ ;\qquad
 Z_l(k)=-{1\over k^2}{\omega(\omega^2-k^2)\over
 (3-\omega^2+k^2)}\Big{|}_{\omega=\omega_l(k)}\ ,\eqno(3.23)
$$
and the cut $\beta_{t,l}$ reads:
$$
 \eqalign{&\beta_t(\omega,k)={3\omega(k^2-\omega^2)\over 4k^3
 \big[(k^2-\omega^2+{3\omega^2\over 2k^2}(1+{k^2-\omega^2\over 2\omega k}
 \ln{k+\omega\over k-\omega}))^2+\big({3\pi\omega\over 4k^3}
 (k^2-\omega^2)\big)^2\big]}\ ;\cr&
 \beta_l(\omega,k)=-{3\omega\over 2k\big[(3+k^2-{3\omega\over 2k}
 \ln{k+\omega\over k-\omega})^2+({3\pi\omega\over 2k})^2\big]}\ .\cr}
 \eqno(3.24)
$$
After we replace the effective propagators and other similar quantities
by their spectral representations, we perform the integrals over the
imaginary times. One such integration yields a delta-function
and the others energy denominators. Actually, we arrange our expressions
in such a way that we have always only two imaginary-time integrals,
which ensures that we get only one energy denominator.
When this is performed, we analytically continue
$ip_0$ to the on-shell real energy $\omega_t(p)+i0^+$. The imaginary part
of the different contributions is obtained using the well known relation
$ 1/(x+i0^+)=\r{Pr}\,1/x-i\pi\delta(x)$. We apply this technique to
the expressions in (3.10), (3.14), (3.17) and (3.20) and sum up the
results to obtain:
$$
 \eqalign{\r{Im}\,\s\Pi_{t3gll}(P)=&{g^2N_cT\over 24\pi}\,
 \int_{0}^{\infty}dk\int_{-\infty}^{+\infty}d\omega_1
 \int_{-\infty}^{+\infty}d\omega_2\,
 \Big[{18k^4\over \omega_1\omega_2}\,\rho_{l1}\,\rho_{l2}
 +{6\omega_1^2\over k\omega_2}\,\Theta_1\,\rho_{l2}
 +{p^2\over 5}\Big[{3k^2\over 2\omega_1\omega_2}\cr&
 \times(3+44k^2-12\omega_1\omega_2)\,\rho_{l1}\,\rho_{l2}
 +{3k\over 2\omega_2}\,\Big(1-{\omega_1^2\over k^2}\Big)\,
 \Big(1-9{\omega_1^2\over k^2}\Big)\,\Theta_1\,\rho_{l2}
 +{4k^3\over \omega_1\omega_2}\,(13+3k^2\cr&
 +10\omega_1-9\omega_1^2)\,
 \rho_{l1}\,\partial_k\rho_{l2}+6{\omega_1\over \omega_2}\,
 \Big(1+2{\omega_1\over k^2}-3{\omega_1^2\over k^2}\Big)\,\Theta_1\,
 \partial_k\rho_{l2}+{2k^4\over\omega_1\omega_2}\,
 (2+5\omega_1)\,\rho_{l1}\,\partial_k^2\rho_{l2}\cr&
 +{3\omega_1^2\over \omega_2k}\,
 \Theta_1\,\partial_k^2\rho_{l2}-{54k^4\over\omega_1\omega_2}\,
 \rho_{l1}\,\rho_{l2}\,\partial_{\omega_1}-{18\omega_1^2\over
 \omega_2k}\Theta_1\,\rho_{l2}\,\partial_{\omega_2}\Big]+\dots\Big]\,
 \delta\big(1-\omega_1-\omega_2\big)\ ,\cr}
 \eqno(3.25)
$$
where $\rho_{li}$ denotes $\rho_l(\omega_i,k)$ with $i=1,2$,
and $\Theta_1\equiv \Theta(k^2-\omega_1^2)$. In (3.25), we have
used $n(\omega_i)\simeq T/\omega_i$ since only soft values of $\omega_i$
are to contribute to the integrals. Also, we note that many terms drop out
in the end because they do not hold an imaginary part. The remaining
integrals in (3.25) are to be carried through numerically, more
on this in the last section.

As we said earlier, the other contributions to $\r{Im}\s\Pi_{t}(P)$ in
(3.7) are manipulated along the same lines. Concerning the $3g$-contributions,
the intermediary steps are much longer, and at times quite intricate.
For example, one thing that doesn't look obvious from the outset is that,
when performing the sum over $k_0$, it is necessary to group the different
terms in a certain manner such that expressions containing $k_0$ in a
way that prevents us from going to the spectral representation 
do cancel out and the technique can then be used.
Another grouping of terms is also necessary in order to ensure
that we end up at each time with only one energy denominator. 
Such subtleties do not occur in the $ll$ contribution we worked out above.
It would be too long and quite cumbersome to report on the intermediary
steps of these contributions, and so we content ourselves with
giving the final resutls, similar to (3.25). For the $lt$ and $tl$
contributions, we find:
$$
 \eqalign{\r{Im}\s\Pi_{t3gtl}(P)=&\,\r{Im}\s\Pi_{t3glt}(P)=
 {g^2N_cT\over 24\pi}\,{1\over 2}\int_0^{\infty}dk
 \int_{-\infty}^{+\infty}{d\omega_1\over \omega_1}
 \int_{-\infty}^{+\infty}{d\omega_2\over \omega_2}\,
 \Big[-18k^2(k^2-\omega_1^2)^2\,\rho_{t1}\,\rho_{l2}\cr&
 -{3\omega_1\over k^3}(k^2-\omega_1^2)^2\,\Theta_1\,\rho_{t2}
 +{6\omega_1\over k}(k^2-\omega_1^2)\,\Theta_1\,\rho_{l2}
 +{p^2\over 5}\Big[\,\Big[{36\over k^2}-20-12k^2+93k^4+50k^6\cr&
 +\Big(-{108\over k^2}+8+12k^2+78k^4\Big)\,\omega_1
 +\Big({72\over k^2}-80-2k^2-186k^4\Big)\,\omega_1^2
 +\Big({72\over k^2}+36-176k^2\Big)\,\omega_1^3\cr&
 +\Big(-{84\over k^2}-179+222k^2\Big)\,\omega_1^4
 +\Big({60\over k^2}+98\Big)\,\omega_1^5
 +\Big({48\over k^2}-86\Big)\,\omega_1^6\,\Big]\,\rho_{t1}\,\rho_{l2}
 +\Big[\Big({3\over 2k}+3k\Big)\omega_1\cr&
 -{3\over k}\omega_1^2
 +\Big({3\over k^3}-{15\over k}\Big)\omega_1^3
 -{9\over k^3}\omega_1^4
 -\Big({57\over 2k^5}-{21\over k^3}\Big)\omega_1^5
 +{12\over k^5}\omega_1^6
 -{9\over k^5}\omega_1^7\,\Big]\,\Theta_1\,\rho_{t2}
 +\Big[-30k\omega_1\cr&
 +{72\over k}\omega_1^3-{42\over k^3}\omega_1^5\,\Big]\,\Theta_1\,\rho_{l2}
 +[-12k-72k^5+(-24k-12k^3+48k^5)\omega_1+(-12k+144k^3)\omega_1^2\cr&
 +(12k-96k^3)\omega_1^3-72k\omega_1^4+48k\omega_1^5\,]\,\rho_{t1}\,
 \partial_k\rho_{l2}
 +[69k+14k^3-3k^5+(-54k+8k^3+6k^5)\omega_1\cr&
 +(2k+2k^3)\omega_1^2+(8k-12k^3)\omega_1^3-11k\omega_1^4
 +6k\omega_1^5\,]\,\rho_{l1}\,\partial_k\rho_{t2}
 +\Big[-{9\over 2}\omega_1+9\omega_1^2+{3\over k^2}\omega_1^3\cr&
 -{18\over k^2}\omega_1^4+{3\over 2k^4}\omega_1^5
 +{9\over k^4}\omega_1^6\,\Big]\,\Theta_1\,\partial_k\rho_{t2}
 +\Big[9\omega_1-18\omega_1^2-{9\over k^2}\omega_1^3
 +{18\over k^2}\omega_1^4\,\Big]\,\Theta_1\,\partial_k\rho_{l2}\cr&
 +[-8k^2-16k^6-16k^2\omega_1+(-8k^2+32k^4)\omega_1^2
 -16k^2\omega_1^4\,]\,\rho_{t1}\,\partial_k^2\rho_{l2}
 +[30k^2+4k^4-2k^6\cr&
 +(-24k^2-8k^4)\omega_1+(-4k^2+4k^4)\omega_1^2
 +8k^2\omega_1^3-2k^2\omega_1^4\,]\,\rho_{l1}\,\partial_k^2\,\rho_{t2}\cr&
 -3{\omega_1\over k^3}(k^2-\omega_1^2)^2
 \,\Theta_1\,\partial_k^2\rho_{t2}
 +6{\omega_1\over k}(k^2-\omega_1^2)\,\Theta_1\,\partial_k^2\rho_{l2}
 +54k^2(k^2-\omega_1^2)^2\,\rho_{t1}\,\rho_{l2}\,\partial_{\omega_1}\cr&
 +9{\omega_1\over k^3}(k^2-\omega_1^2)^2\,\Theta_1
 \,\rho_{t2}\,\partial_{\omega_1}
 -18{\omega_1\over k}(k^2-\omega_1^2)\,\Theta_1\,\rho_{l2}\,
 \partial_{\omega_1}\,\Big]+\dots\,\Big]\,\delta(1-\omega_1-\omega_2)\ .
 \cr}\eqno(3.26)
$$

The above expression is rather long, partly because these two contributions
do not benefit from a possible symmetry between $\omega_1$ and $\omega_2$,
which can lead to some useful simplifications. Indeed, though the algebra
is by far the most tedious for the $tt$ contribution, the use of this
symmetry renders its final result relatively simpler. It reads:
$$
 \eqalign{\r{Im}\s\Pi_{t3gtt}(P)=&{g^2N_cT\over 24\pi}\,\int_0^{\infty}dk
 \int_{-\infty}^{+\infty}{d\omega_1\over \omega_1}
 \int_{-\infty}^{+\infty}{d\omega_2\over \omega_2}
 \Big[36(k^2+\omega_1\omega_2)^2\,\rho_{t1}\,\rho_{t2}
 -6{\omega_1^3\over k^3}(k^2-\omega_1^2)\,\Theta_1\,\rho_{t2}\cr&
 +{p^2\over 5}\Big[\,\Big[-18k^2+122k^4+{1\over 5}(-728+854k^2)\,\omega_1
 \omega_2+\Big({30\over 7k^2}-{706\over 5}\Big)\omega_1^2\omega_2^2
 -{174\over k^2}\omega_1^3\omega_2^3\,\Big]\,\rho_{t1}\,\rho_{t2}\cr&
 +{3\omega_1\over 4k}\Big(1-{\omega_1^2\over k^2}\Big)
 \Big[-5k^2-16\omega_1+2\omega_1^2-
 {21\over k^2}\omega_1^4\Big]\,\Theta_1\,\rho_{t2}
 +\Big[32k^3+20k^5\cr&
 +(112k+124k^3)\,\omega_1+\Big({48\over k}-36k-116k^3\Big)\omega_1^2
 -\Big({108\over k}+232k\Big)\omega_1^3
 +\Big({12\over k}+156k\Big)\omega_1^4\cr&
 +{108\over k}\omega_1^5-{60\over k}\omega_1^6\Big]\,
 \rho_{t1}\,\partial_k\rho_{t2}
 -6{\omega_1^2\over k^2}\Big(1-{\omega_1^2\over k^2}\Big)
 (k^2+3\omega_1-3\omega_1^2)\,\Theta_1\,\partial_k\rho_{t2}\cr&
 +[8k^4+20k^2(1+k^2)\,\omega_1+12(1+k^2)\,\omega_1^2
 -(12+32k^2)\,\omega_1^3-12\omega_1^4+12\omega_1^5\,]\,
 \rho_{t1}\,\partial_k^2\rho_{t2}\cr&
 -3{\omega_1^3\over k^3}(k^2-\omega_1^2)\,\Theta_1\,
 \partial_k^2\rho_{t2}-108(k^2+\omega_1\omega_2)^2\,\rho_{t1}\,
 \rho_{t2}\,\partial_{\omega_1}\cr&
 +18{\omega_1^3\over k^3}(k^2-\omega_1^2)\,\Theta_1\,\rho_{t2}\,
 \partial_{\omega_1}\Big]+\dots\,\Big]\,\delta(1-\omega_1-\omega_2)\ .\cr}
 \eqno(3.27)
$$

The two $4g$-contributions are easier to work out and can be cast in
different ways depending on how we perform the expansion. One compact way
is to write:
$$
 \eqalign{\r{Im}\s\Pi_{t4gl}(P)+\r{Im}\s\Pi_{t4gt}(P)=&
 {g^2N_cT\over 24\pi}\int_0^{\infty}dk
 \int_{-\infty}^{+\infty}d\omega_1\int_{-\infty}^{+\infty}d\omega_2\,
 \Big[-6{k\over \omega_2}\,\Theta_1\,\rho_{l2}
 +{6(k^2-\omega_1^2)\over k\omega_2}\,\Theta_1\,\rho_{t2}\cr&
 +{p^2\over 5}\Big[-{24k^3\over \omega_1\omega_2}\,\delta_1\,\rho_{l2}
 +12{\omega_1\over \omega_2}\,\epsilon(\omega_1)\,\delta_1\,\rho_{t2}
 +{6(4\omega_1-1)\over k\omega_2}\,\Theta_1\,\rho_{t2}\cr&
 +{6k\omega_1\over \omega_2}\,\epsilon(\omega_1)\,\partial_{\omega_1}
 [\delta(k-\omega_1)-\delta(k+\omega_1)\,]\,\rho_{l2}
 +{18k\over \omega_2}\,\Theta_1\,\rho_{l2}\,\partial_{\omega_1}\cr&
 -{18(k^2-\omega_1^2)\over k\omega_2}\,\Theta_1\,\rho_{t2}\,
 \partial_{\omega_1}\,\Big]+\dots\Big]\,\delta(1-\omega_1-\omega_2)
 \ ,\cr}\eqno(3.28)
$$
where $\delta_1$ denotes $\delta(k^2-\omega_1^2)$ and $\epsilon(\omega_1)$
is the sign function. 
\bigskip
\line{\bf 4. Discussion\hfil}
\medskip
Recall that our aim in this work is to calculate to order $p^2$
the damping rate for soft transverse gluons in QCD
at high temparature $T$. We have mentioned in the introduction
that our main motivation is to suggest that the analytic
behavior of this damping rate, and perhaps similar quantities, may
change around some scale $\mu$, see the first section.

The damping rate $\gamma_t(p)$ is given by equation (2.9). 
The denominator is given to order $p^2$ in (2.10). 
The numerator $\r{Im}\s\Pi_t(P)$ is the sum of the expressions given in
(3.25)-(3.28). As we said in the last section, its actual value is to be
determined numerically, an issue we comment on a little later.
But at least from an analytic point of view, our calculation shows
that the expansion in powers of $p^2$ of the damping rate is quite
feasible: we recover the order zero
in $p^2$ given in [2] and find an expression for the next one. Furthermore,
our steps show that the calculation of the higher orders proceeds
in a straightforward, though practically a lot more tedious, manner.

Supposing for a moment that the numerics go smoothly, it is interesting
to ask about the nature of this scale $\mu$. The first suggestion that
comes to mind is that $\mu$ is related to magnetic effects which are
believed to manifest themselves at the next natural order beyond $m_g$,
i.e., $g^2T$. If this is so, there are two interesting points to mention in
this regard. First of all, it would be interesting to understand how
the magnetic effects get into play with respect to other effects and
how and why they make this damping rate and possibly other similar
quantities change their analytic behavior. Also, in the case of
this change being indicative of some sort of phase transition,
it would be interesting to understand the nature of this
transition and the `critical' behavior of QCD around it. 

However, it may be that the scale $\mu$ is due to some other effects. Though
this maybe remote a possibility, it would certainly be very interesting to
contemplate into the matter. One way to start looking into this is to notice
that the coefficients $a_{ti}$ in the expansion (1.1) are all pure numbers.
Thus, with more coefficients calculated and assuming that they are all finite,
it is possible to determine approximately the radius of convergence of the
series, using for exemple a Pad\'e approximant technique, and hence
determining $\mu$ with respect to $m_g$. If by other independent means we
have an expression for the running coupling constant $g$, we can compare
the approximate value of $\mu/m_g$ to this latter and carry a discussion
thereon.

But at this stage, all this is still speculative. Indeed, though, as we said,
the analytic expression of the effective self-energy to order $p^2$ is
quite clean, it is not apriori obvious it is so as far as its numerical
value is concerned.
it is true that the coefficient $a_{t0}$ we get from our calculation
has a finite value, but this is not certain for $a_{t1}$.
First of all, we have expressions involving products of derivatives
of delta-functions we have to clarify precisely how to carry through
with. But more importantly we think is that we are not assured of an
infrared safeness with regard to the integration over the internal soft
momentum $k$. This issue needs careful and thorough investigation, which
is beyond the scope of the present work.

From a physical standpoint,
the fact that $a_{t0}$ is finite means that the quark-gluon plasma is stable
against zero-momentum gluonic excitations. For practical purposes,
it is sufficient to take $\gamma_t(p)\simeq {g^2N_cT\over 24\pi}\,a_{t0}$ for
small enough gluonic momenta. This implicitly presupposes that for such
small non-zero momenta, the plasma remains stable. But this means that, QCD
being the adequate theoretical framework to describe the qgp,
$\gamma_t(p)$ ought to have a smooth analytic behavior for
such small momenta, and hence the higher-than-zero coefficients
in our expansion (1.1) should be expected to be finite. This is the basic
motivation of our work. If on the
contrary they happen to be not so, because of infrared divergences
for instance, then one must look into the matter more closely and try to
reconcile between what we expect and what we get.

After these issues are settled should come the discussion of the physical
implications of these results. For example, the sign of $a_{t0}$ being
positive, is the sign of $a_{t1}$ also positive? If not, then at what
momentum instabilities start to apprear? We think all these issues are worth
pursuing.
\bigskip
\line{\bf Acknowledgements\hfil}
\medskip
A.A. warmly thanks R.E. Shrock for inviting him at Stony Brook where
very early work on this problem has been done. During that stay, he benefited
from stimulating discussions with many people, most particularly
with R.D. Pisarski and I. Zahed. A word of acknowledgement to A.H. Bougourzi
and M. Kacir. He also wants to thank Abdellatif Abada for providing the
original reference that triggered our interest in this work, for his
continued support and warm encouragements. Finally he thanks Asmaa Abada
for caring to provide us with useful references, for her valuable advice
and assistance. O.A. wants to thank O. P\`ene from Orsay for very
stimulating, encouraging and helpful discussions held at a Workshop
in Rabat. She also thanks F. Iddir for all her help.
Finally, we would like to thank A. Kadi-Hanifi for providing
us with some of the references we used and K. Benchallal for spotting
two mistakes in some parts of the calculations.
\vfil\eject
\line{\bf References\hfil}
\medskip
\item{[1]}
K. Geiger, Phys.~Rep.~258 (1995) 237; `{\it Quark Gluon Plasma}', ed.~R.C. Hwa,
World Scientific, 1995; `{\it RHIC Summer Study '97\hskip3pt}', BNL Proceedings;
`{\it QCD and High Energy Hadronic Interactions}',
32$^{\r{nd}}$ Rencontre de Moriond, Proccedings, 1997;
`{\it International School on the
Physics of Quark Gluon Plasma}', Hiroshima, Proceedings
in Prog.~Theor.~Phys., Suppl., 1997; `{\it Quark Matter '96\hs3pt}',
Heidelberg, Proceedings in Nucl.~Phys.~{\bf A} 610 (1996) 1c.

\item{[2]} E. Braaten and R.D. Pisarski,
Phys.~Rev.~{\b D}42 (1990) R2156. Much of the early and recent literature
on the issue of gauge dependance of the damping rate is gathered in
reference 1 of this article.

\item{[3]}
Reference 2; E. Braaten and R.D. Pisarski, Nucl.~Phys.~{\b B}337
(1990) 569; Phys.~Rev.~Lett.~64 (1990) 1338; Nucl.~Phys.~{\b B}339
(1990) 310; J. Frenkel and J.C. Taylor, Nucl.~Phys.~{\bf B}334 (1990) 199;
R.D. Pisarski, Phys.~Rev.~Lett.~63 (1989) 1129;
Physica {\bf A} 158 (1989) 246; Nucl.~Phys.~{\bf A}498 (1989) 423c.

\item{[4]}
R.D. Pisarski, Phys.~Rev.~{\bf D}47 5589 (1993). See also in the context of
QED and with a diiferent approach J.P. Blaizot and E. Iancu,
{\tt hep-ph/9706397}; Phys.~Rev.~{\bf D}55 (1997) 973;
Phys.~Rev.~Lett. 76 (1996) 3080.

\item{[5]}
M.H. Thoma and C.T. Traxler, Phys.~Lett.~{\bf B}378 (1996) 233.

\item{[6]}
R.D. Pisarski, Physica {\bf A} 158 (1989) 146;
Nucl.~Phys.~{\bf B}309 (1988) 476.

\vfill\eject\end